\documentclass[reprint,amssymb,amsmath,aps,pra]{revtex4-1}

\usepackage{graphicx}
\usepackage{dcolumn}
\usepackage{bm}
\usepackage{hyperref}
\usepackage{amsthm}

\newtheorem*{theorem}{Theorem}
\newtheorem*{lemma}{Lemma}
\newtheorem*{corollary}{Corollary}

\newcommand{\la}{\langle}
\newcommand{\ra}{\rangle}

\newcommand{\FS}{{\rm FS}}
\newcommand{\kF}{k_{\rm F}}

\newcommand{\p}{\hat p}
\newcommand{\V}{\hat V}

\newcommand{\mi}{m_{\rm i}}
\newcommand{\mh}{m_{\rm h}}

\newcommand{\be}{\begin{equation}}
\newcommand{\ee}{\end{equation}}

\newcommand{\ket}[1]{\left|{#1}\right\rangle}

\begin{document}


\title{Perpetual motion of a mobile impurity in a one-dimensional quantum gas}

\author{O. Lychkovskiy}
\affiliation{%
 Physics Department, Lancaster University,  Lancaster LA1 4YB, United Kingdom, and
}
\affiliation{Institute for Theoretical and Experimental Physics, 25 B. Cheremushkinskaya,
 Moscow 117218, Russia,
 }
\affiliation{ Russian Quantum Center, Novaya St. 100, Skolkovo, Moscow Region, 143025, Russia.}


\begin{abstract}

Consider an impurity particle injected in a degenerate one-dimensional (1D) gas of noninteracting fermions (or, equivalently, Tonks-Girardeau bosons) with some initial momentum~$p_0$.
We examine the infinite-time value of the momentum of the impurity,~$p_\infty$, as a function of~$p_0$.
A lower bound on~$|p_\infty(p_0)|$ is derived under fairly general conditions. The derivation, based on the existence of the lower edge of the spectrum of the host gas, does not resort to any approximations. The existence of such bound implies the perpetual motion of the impurity in a one-dimensional gas of noninteracting fermions or Tonks-Girardeau bosons at zero temperature. The bound admits an especially simple and useful form when the interaction between the impurity and host particles is everywhere repulsive.

\end{abstract}

\pacs{Valid PACS appear here}
\maketitle


\section{Introduction}
The experimental advances in preparation and manipulation of one-dimensional quantum gases \cite{bloch2008many} has fueled a considerable interest in theoretical studies of their properties. These studies have revealed a variety of remarkable and sometimes unexpected effects \cite{imambekov2012one}. One such recently discovered effect, the absence of complete relaxation of momentum of a mobile impurity injected in a one-dimensional quantum gas, constitutes the subject of the present paper. This phenomenon was predicted in \cite{schecter2012dynamics,schecter2012critical} on the basis of the analysis of the dispersion curve of the impurity-host system. Then the phenomenon was observed in semi-numerical calculations in an integrable model~\cite{mathy2012quantum} and in numerical calculations in families of generically non-integrable models~\cite{mathy2012quantum,knap2014quantum}. These observations were further supported in~\cite{knap2014quantum} by an analytical argument based on the Hellmann-Feynman theorem: it was shown that the average momentum of the impurity in the ground state of an impurity-host system with a fixed total momentum is nonzero. Recently the infinite-time momentum of the impurity for the family of models studied in~\cite{mathy2012quantum} was calculated analytically in the week coupling limit \cite{Burovski2013,Gamayun2014,Gamayun2014keldysh}. The results of the latter calculations confirm the absence of complete  momentum relaxation in both integrable and non-integrable cases. This should be contrasted with the fact that the superfluidity, understood as a dissipationless motion of an infinitely heavy impurity in a quantum liquid, is generally absent in one dimension \cite{astrakharchik2004motion,cherny2012theory}. Moreover, the apparent insensitivity of the above effect to the presence or absence of integrability seems to be at odds with the predicted sharp difference between transport properties of integrable and non-integrable systems \cite{castella1995integrability,zotos1997transport}.

The above conundrums call for further studies of the phenomenon. We aim to contribute to its understanding by rigorously deriving a lower bound on~$|p_\infty(p_0)|$, where $p_0$ and $p_\infty$ are the initial momentum of the impurity and its equilibrium infinite-time momentum, respectively. This bound implies the perpetual motion of the impurity in a one-dimensional gas of noninteracting fermions or Tonks-Girardeau bosons at zero temperature, provided the initial momentum $p_0$ lies in a certain range.

The bound is presented in two forms. The first one is valid for an impurity-host interaction of a general form, however its practical application requires either involving perturbation theory or knowing the exact eigenstates of the impurity-host system. The second one is valid for a narrower class of interactions, namely,  {\it everywhere repulsive} interactions, which are described by a potential $U(x)$ fulfilling the condition
\be\label{potential}
U(x)\geq 0 ~~~~~~ \forall x.
\ee
The advantage of the second bound is that it can be easily calculated for any everywhere repulsive interaction.

\begin{figure}[t]
\includegraphics[width=0.8\linewidth]{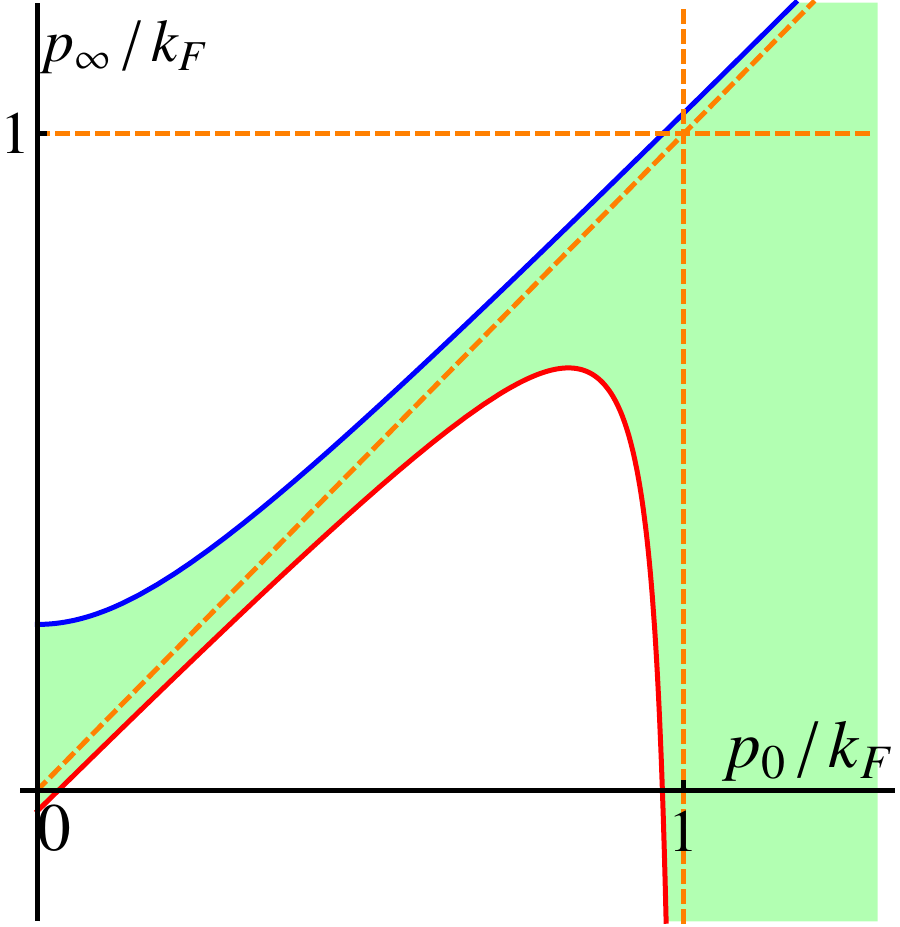}
\caption{(color online) Lower and upper bounds on infinite-time momentum of the mobile impurity, $p_\infty$, for the pointlike interaction (\ref{delta potential}) with $\mi=\mh$ and $g=0.1$,  see eq. (\ref{bounds for delta potential}). Filled region represents the allowed window for $p_\infty$ given a specific $p_0$. For certain initial momenta $p_0$ the infinite-time momentum $p_\infty$ can not be zero.}
\end{figure}

The analytical derivation does not exploit any particular form of interaction (and thus is equally valid both for nonintegrable and integrable systems) and does not resort to any approximations. Thus, our result complements previous works, in which a pointlike interaction was studied and which relied on numerical methods~\cite{mathy2012quantum,knap2014quantum}, on  perturbation theory \cite{Burovski2013,Gamayun2014,Gamayun2014keldysh} or/and on the Bethe anzats solution~\cite{mathy2012quantum,Burovski2013}, the latter being applicable in the integrable case only.

Due to the boson-fermion correspondence valid in one dimension \cite{girardeau1960relationship}, our results are equally applicable for free fermion and Tonks-Girardeau boson hosts. In what follows we stick to a fermion language since it provides more insight into the physics of the problem.


The paper is organized as follows. We introduce the setup in the next section.  The results  are formulated and discussed in Sec. \ref{sec main}. Sec. \ref{sec discussion} is devoted to summary and concluding remarks. The Appendix contains the proofs of the results presented in Sec. \ref{sec main}.

\section{Setup}

Consider a quantum system consisting of $N$ spinless fermions (host particles) and one particle of another species (impurity) in one dimension. The fermions do not interact with each other, but interact with the impurity. The Hamiltonian of the system reads
\be\label{Hamiltonian}
\begin{array}{lcl}
\hat H & = & \hat H^0_{\rm h}+\hat H^0_{\rm i}+\V, \\
\hat H^0_{\rm h} & = & \sum\limits_{n=1}^N\frac{\p_n^2}{2\mh}-C, \\
\hat H^0_{\rm i} & = &\frac{\p_{\rm i}^2}{2\mi}, \\
\V & = & \sum\limits_{n=1}^N U(x_n-x_{\rm i})
\end{array}
\ee
with
\be\label{C}
C \equiv \frac{\kF^2}{6 \mh} \frac{N(N+1)}{N-1}.
\ee
Here  $\mi$ and $\mh$, $\p_{\rm i}$ and $\p_n$, $x_{\rm i}$ and $x_n$ are masses, momenta and coordinates of the impurity and the $n$'th fermion, respectively. Periodic boundary conditions on the wave functions  are imposed, $L$ being the period. The Fermi momentum reads $\kF \equiv (2\pi/L) (N-1)/2,$ and $N$ is assumed to be odd. Constant $C$ assures that the  Fermi sea $|\FS \rangle$, which is the ground state of the free fermion Hamiltonian $\hat H^0_{\rm h}$, has zero kinetic energy:
\be\label{energy of FS}
\hat H^0_{\rm h} |\FS \rangle=0.
\ee

Initially the system is prepared in a product state
\be\label{initial state}
|\FS, p_0\rangle\equiv|\FS \rangle \otimes |p_0\rangle,
\ee
where $|p_0\rangle$ is the plane wave of the impurity with the momentum $p_0$. Without a loss of generality we will always assume $p_0>0$. We are interested in the infinite-time momentum of the impurity defined as
\be\label{pinfty definition}
p_\infty \equiv \lim_{t\rightarrow\infty}\frac1t\int_0^t dt' \langle \FS, p_0|e^{i \hat H t'}\p_{\rm i} e^{-i \hat H t'} | \FS, p_0\rangle.
\ee


\section{\label{sec main}Bounds on infinite-time momentum}
First we provide a simple kinematical argument which suggests that the impurity momentum does not relax whenever $p_0$
is less than certain $q_0$. Consider a pairwise collision between the impurity and one of the fermions from the Fermi sea. In one dimension, final momenta are completely determined by the initial momenta due to the laws of energy and momentum conservation. In addition, the Pauli principle forbids the final momentum of the fermion to be inside the Fermi sea. As a result, the pairwise scattering appears to be impossible for any fermion from the Fermi sea whenever $p_0$ is less than
\be\label{q0}
q_0\equiv \min\{\kF, \frac{\mi}{\mh}\kF\}.
\ee

The above kinematical argument captures the essential physics which lies behind the following
\begin{theorem}
Consider a system of $N$ noninteracting fermions and one impurity particle in one dimension described by the Hamiltonian (\ref{Hamiltonian}), which is prepared in the initial state (\ref{initial state}) with
$$
0<p_0<q_0.
$$
Assume that the eigenvalues of the total Hamiltonian $\hat H$ corresponding to a total momentum $p_0$ are nondegenerate. Then the infinite-time momentum of the impurity, $p_\infty$,
is bounded from below according to
\be\label{lower bound general}
\begin{array}{ll}
p_\infty \geq & p_0 - \frac{\mi}{q_0-p_0}
\left(
\langle \FS, p_0| \V | \FS, p_0\rangle-
\right. \\
& \left.
\sum\limits_{\Psi_E}|\la \FS,p_0|\Psi_E \ra|^2 \langle \Psi_E| \V | \Psi_E \rangle
\right),
\end{array}
\ee
where the sum is over all eigenstates $\Psi_E$ of the total Hamiltonian $\hat H$ with the total momentum $p_0$.
If the interaction between the impurity and the fermions is everywhere repulsive, i.e. the condition  (\ref{potential}) is fulfilled, then a more transparent bound holds:
\be\label{lower bound}
p_\infty \geq p_0 - \frac{\mi\kF \overline U}{\pi(q_0-p_0)} \frac{N}{N-1},
\ee
where $\overline U \equiv\int dx\, U(x)$.
\end{theorem}
This theorem constitutes the main result of the present paper. The proof is provided in the Appendix. Here we only mention that the proof is based on the existence of the lower edge of the spectrum of one-dimensional liquid, the feature which is known to give a prominent impact on physics in one dimension \cite{Giamarchi2003}.

Let us discuss and compare the bounds \eqref{lower bound general} and \eqref{lower bound}. The former has the advantage that it is valid for an arbitrary potential. However, its practical applicability is limited since one has to know exact eigenstates of the total Hamiltonian in order to calculate  the sum in the r.h.s. of  eq. \eqref{lower bound general}. This is generally not feasible, except the integrable cases.  It can be instructive to evaluate the r.h.s. of eq. \eqref{lower bound general} perturbatively. One can immediately see that $O(g)$ terms (where $g$ is a coupling constant) cancel out. Thus, in the week coupling limit the bound \eqref{lower bound general} amounts to $p_0-p_\infty \lesssim O(g^2)$.
It is worth emphasizing at this point that the bound \eqref{lower bound general} itself is derived without resorting to the perturbation theory, however its practical application in nonintegrable cases does require using the perturbation theory.

The assumption \eqref{potential} allows us to drop the eigenstate-dependent term in the r.h.s. of \eqref{lower bound general}, which leads to the bound  \eqref{lower bound} (see Appendix). The latter bound can be easily applied without resorting to any approximations.
The price is twofold. First, the bound \eqref{lower bound} is valid for a narrower class of potentials (which includes, however, an important pointlike repulsive potential studied in \cite{mathy2012quantum,knap2014quantum,Burovski2013}, see below). Second, it is not as tight as the bound  \eqref{lower bound general}, since the difference $p_0-p_\infty$ is bounded by a term linear in the coupling strength.

An additional benefit of the bound \eqref{lower bound} is that it obviates two important points. First, the bound \eqref{lower bound} evidently holds in thermodynamic limit. To take the latter one should merely drop the factor $N/(N-1)$. Second, the range of interaction evidently does not matter: the inequality (\ref{lower bound}) represents a nontrivial bound even for  long range interactions, provided the interaction potential decreases with distance faster than~$1/x$. We expect that both observation generically hold for the bound \eqref{lower bound general} as well.

We complete the discussion of the above theorem by noting that the assumption of nondegeneracy of energy spectrum for a given total momentum is extremely unrestrictive. It is fulfilled in a generic system. Moreover, it is even fulfilled in the integrable case of point-like interaction and equal masses. This can be easily checked by solving Bethe equations for this system~\cite{mcguire1965interacting}.

For completeness, we complement the {\it lower} bound (\ref{lower bound}), valid exclusively in 1D and for $p_0<q_0$, with a rather trivial {\it upper} bound, valid in any dimension and for any $p_0$ (the potential still should be everywhere repulsive):
\be\label{upper bound}
|p_\infty|\leq \sqrt{p_0^2+ \frac{2}{\pi} \mi \kF \overline U \frac{N}{N-1} }.
\ee
This bound is also proven in the Appendix.

Let us consider as an example the case studied in \cite{mathy2012quantum,Burovski2013,Gamayun2014keldysh,Gamayun2014},
\be\label{delta potential}
U(x)=g \delta(x),
\ee
with a positive coupling constant $g$.
In this case, neglecting terms $\sim 1/N$, one obtains from eqs. (\ref{lower bound}) and (\ref{upper bound})
\be\label{bounds for delta potential}
p_0 - \frac{g ~\mi \kF}{\pi(q_0-p_0)}  \leq p_\infty \leq \sqrt{p_0^2+2 g~ \mi \kF/\pi }.
\ee
These bounds are depicted in Fig. 1. One can see that $ p_\infty$ is necessarily nonzero for certain initial conditions. The fact that the impurity is not completely stopped despite the interactions with the host gas is in line with findings of \cite{mathy2012quantum,schecter2012dynamics,schecter2012critical,knap2014quantum,Burovski2013,Gamayun2014, Gamayun2014keldysh}.

\section{\label{sec discussion} Summary and concluding remarks}

To summarise, we have studied the infinite-time momentum of the mobile impurity injected in a degenerate ($T=0$) one-dimensional gas of noninteracting fermions (or, equivalently, Tonks-Girardeau bosons) with some initial momentum~$p_0$. Our main result is represented by the inequalities (\ref{lower bound general}) and (\ref{lower bound}), which bound $|p_\infty(p_0)|$ from below. The bound implies that, contrary to the naive expectation, $p_\infty$ is not zero, at least when $p_0$ lies in a certain range. The derivation of the bounds (\ref{lower bound general}) and (\ref{lower bound}) is completely general and does not rely on perturbation theory, hydrodynamic approximation, exact solution in integrable case or numerical simulations, in contrast to the pioneering papers on the topic \cite{schecter2012dynamics,schecter2012critical,mathy2012quantum,knap2014quantum} and recent papers \cite{Burovski2013,Gamayun2014,Gamayun2014keldysh}. Instead, the derivation rests on the existence of the nontrivial lower edge of the spectrum of the host gas (the same property of the host gas was used to obtain a related analytical result in  \cite{mathy2012quantum}). The only assumption used to derive the bound (\ref{lower bound general}) is the nondegeneracy of the spectrum of the total Hamiltonian for a given total momentum~$p_0$. This assumption is generically fulfilled, even in the presence of integrability. The bound (\ref{lower bound general}) explicitly depends on the exact eigenstates of the total Hamiltonian. Thus in practice it can be applied to the integrable cases or to the cases when the interaction between the impurity and the host can be treated perturbatively. In the latter case the difference $p_0-p_\infty$ (where $p_0>0$) is bounded from above by a $O(g^2)$ term, $g$ being a coupling constant. The dependence of the bound (\ref{lower bound general}) on exact eigenstates precludes, however, its direct application  in nonintegrable cases with interaction strong enough to invalidate the perturbative treatment. An additional assumption that the interaction is everywhere repulsive, as defined in eq.~\eqref{potential}, allows us to overcome this limitation and obtain the bound (\ref{lower bound}), which does not depend on exact eigenstates and can be easily evaluated for an everywhere repulsive potential of arbitrary strength and shape. Remarkably, the inequality (\ref{lower bound}) represents a nontrivial bound even for  long range interactions, provided the interaction potential decreases with distance faster than $1/x$. We also provide a simple upper bound on the infinite time momentum, eq. \eqref{upper bound}, which complements the lower bounds (\ref{lower bound general}) and (\ref{lower bound}).

One may wonder whether the phenomenon proved in the present paper -- the perpetual motion of the impurity -- has anything to do with superfluidity. This is a subtle question. It is known that the superfluidity, understood as a dissipationless motion of an infinitely heavy impurity in a quantum liquid, is absent in one dimension. This can be seen by applying the Landau criterion to the spectrum of excitation of the Luttinger liquid  (see \cite{astrakharchik2004motion} for a more elaborated treatment and review \cite{cherny2012theory} for a clear discussion of various aspects of superfluidity in one dimension). Our result does not contradict this statement, as the r.h.s. of eqs. (\ref{lower bound general}) and (\ref{lower bound})  gives $-\infty$ for $\mi=\infty$. Our bound makes sense  for a mobile impurity of a {\it finite} mass only. Moreover, strictly speaking, it does not guarantee the absence of dissipation -- it only ensures that the impurity is never stopped completely.

The spectrum of excitation has a nontrivial lower edge not only in the free fermion and Tonks-Girardeau gases considered in the present article, but also in generic one-dimensional quantum liquids, as well as in some two- and three-dimensional liquids. For this reason the results obtained here can be generalized for the case of mobile impurity in any such quantum liquid. This generalization will be reported elsewhere.

Let us discus how our findings are related to the results of Ref. \cite{neto1996dynamics}. In the latter work the mobility of a heavy impurity ($\mi\gg \mh$) was calculated at a finite temperature and was found to diverge in the limit of vanishing temperature. This is consistent with our result, however does not by any means automatically imply it. Indeed, since the mobility is defined as $d \overline{v}/dF|_{F=0}$ ($\overline{v}$ being the steady state velocity  of an impurity pulled by a force $F$), it is infinite whenever $\overline{v}(F)\sim F^{\alpha}$ with $\alpha<1$, irrespectively of whether or not the system permits perpetual motion of the impurity in the absence of a force. In agreement with this reasoning, it was found in \cite{Gamayun2014} that $\overline{v}(F)\sim \sqrt{F}$ for a free fermion host with $\mi>\mh$. It should be also noted that the results of \cite{neto1996dynamics} are obtained within the linear response scheme (the force is first considered to be periodic in time, then the limit of vanishing frequency is taken, and {\it after} that the limit of vanishing amplitude is taken), while we work with  a true infinite-time asymptotic state of a conservative impurity-host system. As was emphasised in \cite{neto1996dynamics}, in some cases the results produced by the two approaches can be completely different.

Finally, we briefly discuss our results in the context of thermalization. We have proven that $p_\infty(p_0)$ can not be identically equal to a constant (in particular, to zero), see Fig. 1 for illustration.  This means that the impurity interacting with the bath of one-dimensional noninteracting fermions or Tonks-Girardeau bosons does not thermalize, in particular its equilibrium state depends on the initial condition (see \cite{linden2009quantum} for a thorough analysis of different aspects of thermalization). This is a rare example of thermalization failure in a local, generally non-integrable system without disorder ({\it cf.} discussion in \cite{gogolin2011absence}).
Note, however, that this failure occurs for a special initial state of the bath, namely, the $T=0$ state. We expect that thermalization is restored at $T>0$ in thermodynamic limit.

\begin{acknowledgments}
The author is grateful to V. Cheianov, O. Gamayun, E. Burovskiy, M. Zvonarev and L. Glazman for fruitful discussions. The present work was supported by the ERC grant 279738-NEDFOQ. The author also acknowledges the partial support via grants RFBR-11-02-00441 and RFBR-12-02-00193, the grant for Leading Scientific Schools N$^{\circ} $3830.2014.2, and the Center for Science and Education grant N$^{\circ}$ 8411.
\end{acknowledgments}

\appendix*
\section*{Appendix}
\setcounter{equation}{0}
An important role in the proof of the Theorem is played by the set of eigenstates of the noninteracting Hamiltonian
\be
\hat H^0 \equiv \hat H^0_{\rm h}+\hat H^0_{\rm i}.
\ee
Let us introduce some relevant notations and concepts. Let $K$ be an ordered set $\{k_1,k_2,...,k_N\}$ of fermion momenta, then  $|K\rangle$ is an eigenstate  of $\hat H^0_{\rm h}$ with corresponding momenta, and $E^0_K\equiv\sum_{n=1}^N k_n^2/(2\mh)-C$ is its eigenvalue:
\be
\hat H^0_{\rm h} \ket{K}=E^0_K \ket{K}.
\ee
In this notations the Fermi sea $\ket{\FS}$ corresponds to $K=\{-\kF,-\kF+\delta k,...,\kF\}$, where $\delta k \equiv 2\pi/L.$ The energy of the Fermi sea is zero according to eq. (\ref{energy of FS}), and
\be\label{positivity of energy}
E^0_K\geq E^0_{\FS}=0,
\ee
since $\ket{\FS}$ is the ground state of $\hat H^0_{\rm h}$.
Moreover, a more strong inequality is fulfilled in one dimension, see e.g. \cite{haldane1981luttinger}:
\be\label{edge of spectrum}
E^0_K \geq \frac{|K|(2\kF-|K|)}{2\mh},
\ee
where $|K|\equiv \sum_{n=1}^N k_n$.

Further, let $k$ be a momentum of the impurity, then  $|k\rangle$ is the corresponding plane wave of the impurity:
\be
\hat H^0_{\rm i} \ket{k}=\frac{k^2}{2\mi} \ket{k}.
\ee
Finally, we introduce  product states $\ket{K,k} \equiv \ket{K}\otimes\ket{k}$ which form a complete set of eigenstates of $\hat H^0$:
\be
\hat H^0 \ket{K,k} = E^0_{K,k}\ket{K,k},
\ee
with
\be\label{E0 definition}
E^0_{K,k} \equiv E^0_{K}+\frac{k^2}{2\mi}.
\ee

The total Hamiltonian (\ref{Hamiltonian}) conserves the total momentum which equals~$p_0$ for the state (\ref{initial state}), therefore we will always work in a subspace with the total momentum~$p_0$. This subspace is spanned by all $\ket{K,k}$ with \be\label{total momentum}
|K|+k=p_0.
\ee

The proof of the theorem is based on the following
\begin{lemma}
Consider states $\ket{\FS,p_0}$ and $\ket{K,k}$ with the same total momentum $p_0\in[0,q_0)$. Then
\be\label{key inequality}
p_0-k \leq \frac{\mi}{q_0-p_0} (E^0_{K,k}-E^0_{\FS,p_0}).
\ee
\end{lemma}
{\em Proof.}~
If $k>p_0$, then, using eqs. (\ref{E0 definition}) and (\ref{positivity of energy}) one obtains $E_{K,k}>p_0^2/(2\mi)=E_{\FS,p_0}.$
Thus, the l.h.s. of the inequality (\ref{key inequality}) is negative and the r.h.s. is positive, therefore the inequality (\ref{key inequality}) is fulfilled.

Let us consider now $k\leq p_0$. From eqs. (\ref{E0 definition}), (\ref{edge of spectrum}) and  (\ref{total momentum}) one obtains
\be\label{noname 1}
\frac{E^0_{K,k}-E^0_{\FS,p_0}}{p_0-k} \geq \frac12 \left( \mh^{-1}-\mi^{-1} \right) k +\frac{2\kF-p_0}{2\mh}-\frac{p_0}{2\mi}.
\ee
Consider two cases.\\
{\it (a).} $\mi<\mh$\\
The minimum of the  r.h.s. of eq. (\ref{noname 1}) on the interval $k\in (-\infty, p_0]$  is reached at $k=p_0$ and reads
$\mi^{-1}(\kF \mi/\mh-p_0)$,
which leads to eq. (\ref{key inequality}).\\
{\it (b).} $\mi>\mh$\\
In this case we have to consider intervals $k\in [p_0-2\kF, p_0]$  and $k\in (-\infty,p_0-2\kF]$ separately.\\
{\it (b1).} $k\in [p_0-2\kF, p_0]$\\
The minimum of the  r.h.s. of eq. (\ref{noname 1}) is reached at $k=p_0-2\kF$ and reads $\mi^{-1}(\kF-p_0)$,
which leads to eq.~(\ref{key inequality}).\\
{\it (b2).} $k\in (-\infty,p_0-2\kF]$\\
In this case we do not use eq. (\ref{noname 1}). Instead, we use eqs.~(\ref{E0 definition}) and (\ref{positivity of energy}) to obtain
\be
\frac{E^0_{K,k}-E^0_{\FS,p_0}}{p_0-k} \geq -\frac{k+p_0}{2\mi} \geq \mi^{-1}(\kF-p_0),
\ee
which again leads to eq. (\ref{key inequality}). \\
Note that in the case of equal masses, $\mi=\mh$, the result (\ref{key inequality}) is obtained directly from eq. (\ref{noname 1}).$\blacksquare$

\bigskip

\begin{corollary}
State $\ket{\FS,p_0}$ is a ground state of $\hat H^0$ in the subspace with the total momentum $p_0$.
\end{corollary}
{\em Proof.} If $k>p_0$, the proof is contained in the proof of the Lemma. If $k\leq p_0$, the Corollary follows from the inequality (\ref{key inequality}). $\blacksquare$

\bigskip

{\em Proof of the Theorem.}
Let us expand the evolution operator in  the definition (\ref{pinfty definition}) as $e^{-i\hat H t'}= \sum_{\Psi_E} |\Psi_E \ra \la \Psi_E|e^{- iE t'}$. Then, due  the assumed non-degeneracy of eigenvalues, only the diagonal in energy terms survive the integration in the
$t \rightarrow \infty$ limit, and one is left with
\be
p_\infty = \sum_{\Psi_E}|\la \FS,p_0|\Psi_E \ra|^2 \la \Psi_E|\p_{\rm i}|\Psi_E\ra.
\ee
Further,
\begin{align}
p_0-p_\infty & =\sum_{\Psi_E}\sum_{K}\sum_{k=-\infty}^{+\infty} (p_0-k) |\la \FS,p_0|\Psi_E \ra|^2 |\la \Psi_E|K,k\ra|^2.
\end{align}
Here the summations are performed over the eigenstates $\ket{\Psi_E}$ of the total Hamiltonian $\hat H$, and over the eigenstates   $\ket {K,k}$ of the nonimteracting Hamiltonian~$\hat H^0$.

The key step is to implement the inequality (\ref{key inequality}) which is valid for {\it any} $K$:
\begin{align}\label{noname 2}
p_0-p_\infty  \leq & \frac{ \mi }{q_0-p_0} \sum\limits_{\Psi_E}|\la \FS,p_0|\Psi_E \ra|^2 \times  \nonumber\\
                    & \times \sum\limits_{K}\sum\limits_{k=-\infty}^{\infty}
 \la \Psi_E|\hat H^0-E^0_{\FS,p_0}|K,k\ra \la K,k|\Psi_E\ra \nonumber\\
 = &  \frac{ \mi }{q_0-p_0} \sum\limits_{\Psi_E}|\la \FS,p_0|\Psi_E \ra|^2  \la \Psi_E|\hat H^0-E^0_{\FS,p_0} | \Psi_E\ra.
\end{align}

Taking into account that
\be
\sum_{\Psi_E}  |\la \FS,p_0|\Psi_E \ra|^2  (E-E^0_{\FS,p_0}) =\la \FS,p_0| \V | \FS,p_0\ra,
\ee
one finally obtains the desired bound (\ref{lower bound general}).

To obtain the bound \eqref{lower bound} from the bound \eqref{lower bound general}, one has to drop the second term in the brackets in the r.h.s. of \eqref{lower bound general} taking into account that according to the definition~\eqref{potential}
\be\label{positivity of V}
\la\Psi|\V|\Psi\ra \geq 0~~~~~\forall~ \Psi
\ee for everywhere repulsive interactions, and to use
\be
\la \FS,p_0| \V | \FS,p_0\ra=\frac{\kF \overline U}{\pi}\frac{N}{N-1}.
\ee
$\blacksquare$

\bigskip
{\em Proof of the upper bound (\ref{upper bound}).}
First let us derive a bound on
\be
(p^2)_\infty \equiv \lim_{t\rightarrow\infty}\frac1t\int_0^t dt' \langle \FS, p_0|e^{i \hat H t'}\p_{\rm i}^2 e^{-i \hat H t'} | \FS, p_0\rangle.
\ee
If the spectrum of $\hat H$ is nondegenerate for a given total momentum $p_0$, then
\begin{align}\label{noname 5}
& (p^2)_\infty=\sum_{\Psi_E}|\la \FS,p_0|\Psi_E \ra|^2 \la \Psi_E|\p_{\rm i}^2|\Psi_E\ra
\nonumber\\
& =2\mi\sum_{\Psi_E}|\la \FS,p_0|\Psi_E \ra|^2 (E-\la\Psi_E|\hat H^0_{\rm h}|\Psi_E\ra-\la\Psi_E|\V |\Psi_E\ra)
\nonumber\\
& \leq 2\mi \la \FS,p_0| \hat H | \FS,p_0\ra=p_0^2+ 2\mi \la \FS,p_0| \V | \FS,p_0\ra.
\nonumber\\
\end{align}
The inequality is obtained with the use of eqs. (\ref{positivity of energy}) and (\ref{positivity of V}).
Since $(p_\infty)^2 \leq (p^2)_\infty$, the bound (\ref{upper bound}) is a direct consequence of eq. (\ref{noname 5}). $\blacksquare$
\bibliography{C:/D/Work/QM/Bibs/1D}

\providecommand{\noopsort}[1]{}\providecommand{\singleletter}[1]{#1}%
\begin{thebibliography}{20}%
\makeatletter
\providecommand \@ifxundefined [1]{%
 \@ifx{#1\undefined}
}%
\providecommand \@ifnum [1]{%
 \ifnum #1\expandafter \@firstoftwo
 \else \expandafter \@secondoftwo
 \fi
}%
\providecommand \@ifx [1]{%
 \ifx #1\expandafter \@firstoftwo
 \else \expandafter \@secondoftwo
 \fi
}%
\providecommand \natexlab [1]{#1}%
\providecommand \enquote  [1]{``#1''}%
\providecommand \bibnamefont  [1]{#1}%
\providecommand \bibfnamefont [1]{#1}%
\providecommand \citenamefont [1]{#1}%
\providecommand \href@noop [0]{\@secondoftwo}%
\providecommand \href [0]{\begingroup \@sanitize@url \@href}%
\providecommand \@href[1]{\@@startlink{#1}\@@href}%
\providecommand \@@href[1]{\endgroup#1\@@endlink}%
\providecommand \@sanitize@url [0]{\catcode `\\12\catcode `\$12\catcode
  `\&12\catcode `\#12\catcode `\^12\catcode `\_12\catcode `\%12\relax}%
\providecommand \@@startlink[1]{}%
\providecommand \@@endlink[0]{}%
\providecommand \url  [0]{\begingroup\@sanitize@url \@url }%
\providecommand \@url [1]{\endgroup\@href {#1}{\urlprefix }}%
\providecommand \urlprefix  [0]{URL }%
\providecommand \Eprint [0]{\href }%
\providecommand \doibase [0]{http://dx.doi.org/}%
\providecommand \selectlanguage [0]{\@gobble}%
\providecommand \bibinfo  [0]{\@secondoftwo}%
\providecommand \bibfield  [0]{\@secondoftwo}%
\providecommand \translation [1]{[#1]}%
\providecommand \BibitemOpen [0]{}%
\providecommand \bibitemStop [0]{}%
\providecommand \bibitemNoStop [0]{.\EOS\space}%
\providecommand \EOS [0]{\spacefactor3000\relax}%
\providecommand \BibitemShut  [1]{\csname bibitem#1\endcsname}%
\let\auto@bib@innerbib\@empty
\bibitem [{\citenamefont {Bloch}\ \emph {et~al.}(2008)\citenamefont {Bloch},
  \citenamefont {Dalibard},\ and\ \citenamefont {Zwerger}}]{bloch2008many}%
  \BibitemOpen
  \bibfield  {author} {\bibinfo {author} {\bibfnamefont {I.}~\bibnamefont
  {Bloch}}, \bibinfo {author} {\bibfnamefont {J.}~\bibnamefont {Dalibard}}, \
  and\ \bibinfo {author} {\bibfnamefont {W.}~\bibnamefont {Zwerger}},\
  }\href@noop {} {\bibfield  {journal} {\bibinfo  {journal} {Reviews of Modern
  Physics}\ }\textbf {\bibinfo {volume} {80}},\ \bibinfo {pages} {885}
  (\bibinfo {year} {2008})}\BibitemShut {NoStop}%
\bibitem [{\citenamefont {Imambekov}\ \emph {et~al.}(2012)\citenamefont
  {Imambekov}, \citenamefont {Schmidt},\ and\ \citenamefont
  {Glazman}}]{imambekov2012one}%
  \BibitemOpen
  \bibfield  {author} {\bibinfo {author} {\bibfnamefont {A.}~\bibnamefont
  {Imambekov}}, \bibinfo {author} {\bibfnamefont {T.~L.}\ \bibnamefont
  {Schmidt}}, \ and\ \bibinfo {author} {\bibfnamefont {L.~I.}\ \bibnamefont
  {Glazman}},\ }\href {\doibase 10.1103/RevModPhys.84.1253} {\bibfield
  {journal} {\bibinfo  {journal} {Rev. Mod. Phys.}\ }\textbf {\bibinfo {volume}
  {84}},\ \bibinfo {pages} {1253} (\bibinfo {year} {2012})}\BibitemShut
  {NoStop}%
\bibitem [{\citenamefont {Schecter}\ \emph
  {et~al.}(2012{\natexlab{a}})\citenamefont {Schecter}, \citenamefont
  {Gangardt},\ and\ \citenamefont {Kamenev}}]{schecter2012dynamics}%
  \BibitemOpen
  \bibfield  {author} {\bibinfo {author} {\bibfnamefont {M.}~\bibnamefont
  {Schecter}}, \bibinfo {author} {\bibfnamefont {D.}~\bibnamefont {Gangardt}},
  \ and\ \bibinfo {author} {\bibfnamefont {A.}~\bibnamefont {Kamenev}},\
  }\href@noop {} {\bibfield  {journal} {\bibinfo  {journal} {Annals of
  Physics}\ }\textbf {\bibinfo {volume} {327}},\ \bibinfo {pages} {639}
  (\bibinfo {year} {2012}{\natexlab{a}})}\BibitemShut {NoStop}%
\bibitem [{\citenamefont {Schecter}\ \emph
  {et~al.}(2012{\natexlab{b}})\citenamefont {Schecter}, \citenamefont
  {Kamenev}, \citenamefont {Gangardt},\ and\ \citenamefont
  {Lamacraft}}]{schecter2012critical}%
  \BibitemOpen
  \bibfield  {author} {\bibinfo {author} {\bibfnamefont {M.}~\bibnamefont
  {Schecter}}, \bibinfo {author} {\bibfnamefont {A.}~\bibnamefont {Kamenev}},
  \bibinfo {author} {\bibfnamefont {D.~M.}\ \bibnamefont {Gangardt}}, \ and\
  \bibinfo {author} {\bibfnamefont {A.}~\bibnamefont {Lamacraft}},\ }\href
  {\doibase 10.1103/PhysRevLett.108.207001} {\bibfield  {journal} {\bibinfo
  {journal} {Phys. Rev. Lett.}\ }\textbf {\bibinfo {volume} {108}},\ \bibinfo
  {pages} {207001} (\bibinfo {year} {2012}{\natexlab{b}})}\BibitemShut
  {NoStop}%
\bibitem [{\citenamefont {Mathy}\ \emph {et~al.}(2012)\citenamefont {Mathy},
  \citenamefont {Zvonarev},\ and\ \citenamefont {Demler}}]{mathy2012quantum}%
  \BibitemOpen
  \bibfield  {author} {\bibinfo {author} {\bibfnamefont {C.~J.}\ \bibnamefont
  {Mathy}}, \bibinfo {author} {\bibfnamefont {M.~B.}\ \bibnamefont {Zvonarev}},
  \ and\ \bibinfo {author} {\bibfnamefont {E.}~\bibnamefont {Demler}},\
  }\href@noop {} {\bibfield  {journal} {\bibinfo  {journal} {Nature Physics}\
  }\textbf {\bibinfo {volume} {8}},\ \bibinfo {pages} {881} (\bibinfo {year}
  {2012})}\BibitemShut {NoStop}%
\bibitem [{\citenamefont {Knap}\ \emph {et~al.}(2014)\citenamefont {Knap},
  \citenamefont {Mathy}, \citenamefont {Ganahl}, \citenamefont {Zvonarev},\
  and\ \citenamefont {Demler}}]{knap2014quantum}%
  \BibitemOpen
  \bibfield  {author} {\bibinfo {author} {\bibfnamefont {M.}~\bibnamefont
  {Knap}}, \bibinfo {author} {\bibfnamefont {C.~J.}\ \bibnamefont {Mathy}},
  \bibinfo {author} {\bibfnamefont {M.}~\bibnamefont {Ganahl}}, \bibinfo
  {author} {\bibfnamefont {M.~B.}\ \bibnamefont {Zvonarev}}, \ and\ \bibinfo
  {author} {\bibfnamefont {E.}~\bibnamefont {Demler}},\ }\href@noop {}
  {\bibfield  {journal} {\bibinfo  {journal} {Physical Review Letters}\
  }\textbf {\bibinfo {volume} {112}},\ \bibinfo {pages} {015302} (\bibinfo
  {year} {2014})}\BibitemShut {NoStop}%
\bibitem [{\citenamefont {Burovski}\ \emph {et~al.}(2013)\citenamefont
  {Burovski}, \citenamefont {Cheianov}, \citenamefont {Gamayun},\ and\
  \citenamefont {Lychkovskiy}}]{Burovski2013}%
  \BibitemOpen
  \bibfield  {author} {\bibinfo {author} {\bibfnamefont {E.}~\bibnamefont
  {Burovski}}, \bibinfo {author} {\bibfnamefont {V.}~\bibnamefont {Cheianov}},
  \bibinfo {author} {\bibfnamefont {O.}~\bibnamefont {Gamayun}}, \ and\
  \bibinfo {author} {\bibfnamefont {O.}~\bibnamefont {Lychkovskiy}},\
  }\href@noop {} {\bibfield  {journal} {\bibinfo  {journal} {arXiv:1308.6147}\
  } (\bibinfo {year} {2013})}\BibitemShut {NoStop}%
\bibitem [{\citenamefont {Gamayun}\ \emph {et~al.}(2014)\citenamefont
  {Gamayun}, \citenamefont {Lychkovskiy},\ and\ \citenamefont
  {Cheianov}}]{Gamayun2014}%
  \BibitemOpen
  \bibfield  {author} {\bibinfo {author} {\bibfnamefont {O.}~\bibnamefont
  {Gamayun}}, \bibinfo {author} {\bibfnamefont {O.}~\bibnamefont
  {Lychkovskiy}}, \ and\ \bibinfo {author} {\bibfnamefont {V.}~\bibnamefont
  {Cheianov}},\ }\href@noop {} {\bibfield  {journal} {\bibinfo  {journal}
  {arXiv:1402.6362}\ } (\bibinfo {year} {2014})}\BibitemShut {NoStop}%
\bibitem [{\citenamefont {Gamayun}(2014)}]{Gamayun2014keldysh}%
  \BibitemOpen
  \bibfield  {author} {\bibinfo {author} {\bibfnamefont {O.}~\bibnamefont
  {Gamayun}},\ }\href@noop {} {\bibfield  {journal} {\bibinfo  {journal}
  {arXiv:1402.7064}\ } (\bibinfo {year} {2014})}\BibitemShut {NoStop}%
\bibitem [{\citenamefont {Astrakharchik}\ and\ \citenamefont
  {Pitaevskii}(2004)}]{astrakharchik2004motion}%
  \BibitemOpen
  \bibfield  {author} {\bibinfo {author} {\bibfnamefont {G.}~\bibnamefont
  {Astrakharchik}}\ and\ \bibinfo {author} {\bibfnamefont {L.}~\bibnamefont
  {Pitaevskii}},\ }\href@noop {} {\bibfield  {journal} {\bibinfo  {journal}
  {Physical Review A}\ }\textbf {\bibinfo {volume} {70}},\ \bibinfo {pages}
  {013608} (\bibinfo {year} {2004})}\BibitemShut {NoStop}%
\bibitem [{\citenamefont {Cherny}\ \emph {et~al.}(2012)\citenamefont {Cherny},
  \citenamefont {Caux},\ and\ \citenamefont {Brand}}]{cherny2012theory}%
  \BibitemOpen
  \bibfield  {author} {\bibinfo {author} {\bibfnamefont {A.~Y.}\ \bibnamefont
  {Cherny}}, \bibinfo {author} {\bibfnamefont {J.-S.}\ \bibnamefont {Caux}}, \
  and\ \bibinfo {author} {\bibfnamefont {J.}~\bibnamefont {Brand}},\
  }\href@noop {} {\bibfield  {journal} {\bibinfo  {journal} {Frontiers of
  Physics}\ }\textbf {\bibinfo {volume} {7}},\ \bibinfo {pages} {54} (\bibinfo
  {year} {2012})}\BibitemShut {NoStop}%
\bibitem [{\citenamefont {Castella}\ \emph {et~al.}(1995)\citenamefont
  {Castella}, \citenamefont {Zotos},\ and\ \citenamefont
  {Prelov{\v{s}}ek}}]{castella1995integrability}%
  \BibitemOpen
  \bibfield  {author} {\bibinfo {author} {\bibfnamefont {H.}~\bibnamefont
  {Castella}}, \bibinfo {author} {\bibfnamefont {X.}~\bibnamefont {Zotos}}, \
  and\ \bibinfo {author} {\bibfnamefont {P.}~\bibnamefont {Prelov{\v{s}}ek}},\
  }\href@noop {} {\bibfield  {journal} {\bibinfo  {journal} {Physical review
  letters}\ }\textbf {\bibinfo {volume} {74}},\ \bibinfo {pages} {972}
  (\bibinfo {year} {1995})}\BibitemShut {NoStop}%
\bibitem [{\citenamefont {Zotos}\ \emph {et~al.}(1997)\citenamefont {Zotos},
  \citenamefont {Naef},\ and\ \citenamefont {Prelovsek}}]{zotos1997transport}%
  \BibitemOpen
  \bibfield  {author} {\bibinfo {author} {\bibfnamefont {X.}~\bibnamefont
  {Zotos}}, \bibinfo {author} {\bibfnamefont {F.}~\bibnamefont {Naef}}, \ and\
  \bibinfo {author} {\bibfnamefont {P.}~\bibnamefont {Prelovsek}},\ }\href@noop
  {} {\bibfield  {journal} {\bibinfo  {journal} {Physical Review B}\ }\textbf
  {\bibinfo {volume} {55}},\ \bibinfo {pages} {11029} (\bibinfo {year}
  {1997})}\BibitemShut {NoStop}%
\bibitem [{\citenamefont {Girardeau}(1960)}]{girardeau1960relationship}%
  \BibitemOpen
  \bibfield  {author} {\bibinfo {author} {\bibfnamefont {M.}~\bibnamefont
  {Girardeau}},\ }\href@noop {} {\bibfield  {journal} {\bibinfo  {journal}
  {Journal of Mathematical Physics}\ }\textbf {\bibinfo {volume} {1}},\
  \bibinfo {pages} {516} (\bibinfo {year} {1960})}\BibitemShut {NoStop}%
\bibitem [{\citenamefont {Giamarchi}(2003)}]{Giamarchi2003}%
  \BibitemOpen
  \bibfield  {author} {\bibinfo {author} {\bibfnamefont {T.}~\bibnamefont
  {Giamarchi}},\ }\href@noop {} {\emph {\bibinfo {title} {Quantum Physics in
  One Dimension}}}\ (\bibinfo  {publisher} {Oxford University Press},\ \bibinfo
  {year} {2003})\BibitemShut {NoStop}%
\bibitem [{\citenamefont {McGuire}(1965)}]{mcguire1965interacting}%
  \BibitemOpen
  \bibfield  {author} {\bibinfo {author} {\bibfnamefont {J.}~\bibnamefont
  {McGuire}},\ }\href@noop {} {\bibfield  {journal} {\bibinfo  {journal}
  {Journal of Mathematical Physics}\ }\textbf {\bibinfo {volume} {6}},\
  \bibinfo {pages} {432} (\bibinfo {year} {1965})}\BibitemShut {NoStop}%
\bibitem [{\citenamefont {Neto}\ and\ \citenamefont
  {Fisher}(1996)}]{neto1996dynamics}%
  \BibitemOpen
  \bibfield  {author} {\bibinfo {author} {\bibfnamefont {A.~C.}\ \bibnamefont
  {Neto}}\ and\ \bibinfo {author} {\bibfnamefont {M.~P.}\ \bibnamefont
  {Fisher}},\ }\href@noop {} {\bibfield  {journal} {\bibinfo  {journal}
  {Physical Review B}\ }\textbf {\bibinfo {volume} {53}},\ \bibinfo {pages}
  {9713} (\bibinfo {year} {1996})}\BibitemShut {NoStop}%
\bibitem [{\citenamefont {Linden}\ \emph {et~al.}(2009)\citenamefont {Linden},
  \citenamefont {Popescu}, \citenamefont {Short},\ and\ \citenamefont
  {Winter}}]{linden2009quantum}%
  \BibitemOpen
  \bibfield  {author} {\bibinfo {author} {\bibfnamefont {N.}~\bibnamefont
  {Linden}}, \bibinfo {author} {\bibfnamefont {S.}~\bibnamefont {Popescu}},
  \bibinfo {author} {\bibfnamefont {A.~J.}\ \bibnamefont {Short}}, \ and\
  \bibinfo {author} {\bibfnamefont {A.}~\bibnamefont {Winter}},\ }\href@noop {}
  {\bibfield  {journal} {\bibinfo  {journal} {Physical Review E}\ }\textbf
  {\bibinfo {volume} {79}},\ \bibinfo {pages} {061103} (\bibinfo {year}
  {2009})}\BibitemShut {NoStop}%
\bibitem [{\citenamefont {Gogolin}\ \emph {et~al.}(2011)\citenamefont
  {Gogolin}, \citenamefont {M{\"u}ller},\ and\ \citenamefont
  {Eisert}}]{gogolin2011absence}%
  \BibitemOpen
  \bibfield  {author} {\bibinfo {author} {\bibfnamefont {C.}~\bibnamefont
  {Gogolin}}, \bibinfo {author} {\bibfnamefont {M.~P.}\ \bibnamefont
  {M{\"u}ller}}, \ and\ \bibinfo {author} {\bibfnamefont {J.}~\bibnamefont
  {Eisert}},\ }\href@noop {} {\bibfield  {journal} {\bibinfo  {journal}
  {Physical Review Letters}\ }\textbf {\bibinfo {volume} {106}},\ \bibinfo
  {pages} {040401} (\bibinfo {year} {2011})}\BibitemShut {NoStop}%
\bibitem [{\citenamefont {Haldane}(1981)}]{haldane1981luttinger}%
  \BibitemOpen
  \bibfield  {author} {\bibinfo {author} {\bibfnamefont {F.}~\bibnamefont
  {Haldane}},\ }\href@noop {} {\bibfield  {journal} {\bibinfo  {journal}
  {Journal of Physics C: Solid State Physics}\ }\textbf {\bibinfo {volume}
  {14}},\ \bibinfo {pages} {2585} (\bibinfo {year} {1981})}\BibitemShut
  {NoStop}%
\end{thebibliography}%

\end{document}